# Scattering-Free Optical Edge States between Heterogeneous Photonic Topological Insulators

Tzuhsuan Ma and Gennady Shvets*

Department of Physics, The University of Texas at Austin, Austin, Texas 78712, USA

*gena@physics.utexas.edu

We propose a set of three simple photonic platforms capable of emulating quantum topologically insulating phases corresponding to Hall, spin-Hall, and valley-Hall effects. It is shown that an interface between any two of these heterogeneous photonic topological insulators supports scattering-free surface states. Spin and valley degrees of freedom characterizing such topologically protected surface waves determine their unique pathways through complex photonic circuits comprised of multiple heterogeneous interfaces.

Light propagation through waveguides, photonic crystals, and other photonic system can often be reduced to a simple scalar wave equation that imposes fundamental limitations on how optical energy can be transported in space. For example, it is generally believed to be impossible to guide light along sharply bent trajectories without reflections. A new paradigm that calls into question this conventional wisdom has been recently introduced with the realization of a new class of photonic structures: photonic topological insulators (PTIs) [ 1, 2, 3, 4, 5, 6, 7, 8, 9]. Just as their condensed matter counterparts, topological insulators (TIs) [ 10, 11, 12] from which they have been derived by analogy, PTIs enable reflections-free propagation of topologically protected surfaces waves (TPSWs) [ 5, 8] along almost arbitrarily shaped interfaces separating two different PTIs. For example, several photonic analogs of a quantum spin-Hall (QSH) [ 4, 5, 8] topological insulators have been recently proposed. However, the library of known condensed materials supporting topological electronic states extends much beyond QSH TIs, including the quantum Hall (QH) [ 13] and quantum-valley Hall (QVH) [ 14, 15, 16, 17] topological insulators.

In this Letter, we demonstrate that these three condensed matter systems can be emulated by novel PTI structures. More significantly, while it is nearly impossible to realize lateral heterojunctions between different classes (e.g., between QSH and QH) of TIs in naturally occurring electronic materials, interfacing heterogeneous PTIs is relatively straightforward and, in fact, highly beneficial to developing novel photonic circuits. We demonstrate that reflections-free TPSWs are supported by the interfaces between such domains. It is shown that topological protection emerges not only from the conservation of the spin degree of freedom (DOF) [ 4, 5, 8], but also from the conservation of another binary DOF: the *valley*. While it has been recently recognized that the valley DOF can produce novel topological phases and chiral edge states of electrons [ 18, 19, 20], here we demonstrate its significance to reflections-free propagation of surface waves in photonic structures.

The photonic analogues of the three topological phases are obtained by imposing three types of distinct symmetry-breaking perturbations on a simple symmetric "photonic graphene" (PhG) [ 21] structure shown in Fig.1(a). The PhG is comprised of a hexagonal arrays of metal posts symmetrically placed between two metal plates and separated from them by the gap $g_0$. A similar photonic structure was considered earlier in the narrower context of QSH-type [ 8] PTIs that can be obtained from the PhG by the first, anisotropy-inducing, perturbation shown in Fig.2(a) which preserves both the time-reversal ($T$) and the in-plane parity ($P$) symmetries. The second perturbation of the unit cell, which breaks the $T$-symmetry but preserves the $P$-symmetry, involves inserting gyromagnetic material into the gaps as shown in Fig.2(b). Under an external magnetic field $B_z$, it acquires off-diagonal elements of the permeability tensor: $\Delta\mu_{yx} = -\Delta\mu_{xy} \equiv \kappa \propto B_z$. Finally, the third perturbation of the rod's shape shown in Fig.2(c) breaks the $P$-symmetry while preserving the $T$-symmetry. As demonstrated below, these perturbations give rise to distinct types of PTIs that emulate the QSH, QH, and QVH effects, respectively, and which can be combined to create domain walls supporting new types of TPSWs.

The lowest-order mutually orthogonal eigenmodes of the PhG can be classified as TM- and TE-like based on their field profiles shown in Fig.1(b). Because only the lowest order TE and TM modes exist in the $K$ and $K'$ valleys, any electromagnetic mode with the Bloch wavenumber $\boldsymbol{k}_\perp$ can be expanded as

$$\boldsymbol{E}(\boldsymbol{r},t) = \sum_{n,\boldsymbol{k}_\perp}\left[a_e^n(\boldsymbol{k}_\perp)\boldsymbol{e}_e^{n,\boldsymbol{k}_\perp}(\boldsymbol{r}_\perp,z) + a_m^n(\boldsymbol{k}_\perp)\boldsymbol{e}_m^{n,\boldsymbol{k}_\perp}(\boldsymbol{r}_\perp,z)\right]e^{i\boldsymbol{k}_\perp\cdot\boldsymbol{r}_\perp - i\omega_n(\boldsymbol{k}_\perp)t} + c.c. \quad (1)$$

$$\boldsymbol{H}(\boldsymbol{r},t) = \sum_{n,\boldsymbol{k}_\perp}\left[a_e^n(\boldsymbol{k}_\perp)\boldsymbol{h}_e^{n,\boldsymbol{k}_\perp}(\boldsymbol{r}_\perp,z) + a_m^n(\boldsymbol{k}_\perp)\boldsymbol{h}_m^{n,\boldsymbol{k}_\perp}(\boldsymbol{r}_\perp,z)\right]e^{i\boldsymbol{k}_\perp\cdot\boldsymbol{r}_\perp - i\omega_n(\boldsymbol{k}_\perp)t} + c.c., \quad (2)$$

where the $n = 1,2$ index refers to lower (upper) propagation bands, and $\boldsymbol{e}_e^{n,\boldsymbol{k}_\perp}$, $\boldsymbol{h}_e^{n,\boldsymbol{k}_\perp}$, $\boldsymbol{e}_m^{n,\boldsymbol{k}_\perp}$ and $\boldsymbol{h}_m^{n,\boldsymbol{k}_\perp}$ are the normalized field profiles chosen to be periodic in the $\boldsymbol{r}_\perp = (x,y)$ plane. The TE/TM-like modes, labelled as $e/m$, are defined by the parity of $\hat{\boldsymbol{z}} \cdot \boldsymbol{h}_{e/m}$ with respect $z$ as illustrated in Fig.1(b). The eigenfrequencies $\omega_n(\boldsymbol{k}_\perp)$ of the two modes were calculated as functions of the Bloch wavenumber $\boldsymbol{k}_\perp = (k_x, k_y)$ inside the Brilloine zone (BZ) using COMSOL Multiphysics code for a specific PhG with a lattice constant $a_0$, inter-plate distance $h_0$, cylinders' diameter $d_0$, and the gap size $g_0$ (see caption for their values).

The resulting dispersion curves are plotted in Fig.1(c) as dashed (TM) and dotted (TE) lines along with the higher-order (solid lines) modes neglected in this study. The $C_{3v}$ wave vector symmetry group [22] results in doubly degenerate modes at the $K/K'$ edges of the BZ shown in Fig.1(c). Each mode forms its own Dirac cone in the non-equivalent "valleys" [14, 15, 16] of $K$ and $K'$ centered at the respective Dirac frequency $\omega_D^{e,m}$. This degeneracy enables the choice of eigenmodes that have the right- or left-hand circular polarizations (RCP and LCP) in the mid-plane ($z = 0$), thereby imparting both TE and TM waves with an orbital degree of freedom. Therefore, at the $K/K'$ edges we choose $n = 1,2$ in Eq.(1) to represent the LCP and RCP orbital states, respectively. Moreover, by a judicious choice of $h_0$, $d_0$, and $g_0$, it is possible to achieve the inter-mode degeneracy [4, 8], i.e. $\omega_D^e = \omega_D^m \equiv \omega_D$ and $\partial\omega_D^e/\partial k = \partial\omega_D^m/\partial k \equiv v_D$ as shown in Fig.1(c).

For simplicity, we first concentrate on the $K$-valley of the BZ, where a finite $\delta\boldsymbol{k} \equiv \boldsymbol{k}_\perp - \boldsymbol{K}$ (where $\boldsymbol{K} = \boldsymbol{e}_x 4\pi/3a_0$) lifts the Dirac degeneracy. The effective $K$-valley Hamiltonian [8] expressed in the RCP/LCP (orbital) basis is $\mathcal{H}_{0K}^{e,m}(\delta\boldsymbol{k}) = v_D(\delta k_x\hat{\sigma}_x + \delta k_y\hat{\sigma}_y)$, where $\hat{\sigma}_{x,y,z}$ are the Pauli matrices acting on the orbital state vector $\mathbf{U}_K^{e,m} = [a_{e,m}^R; a_{e,m}^L]$. The degenerate expansion basis $\mathbf{U}_R = [1; 0]$ and $\mathbf{U}_L = [0; 1]$ is defined according to its transformation with respect to the $2\pi/3$ rotation $\mathcal{R}_3$, according to $\mathcal{R}_3 \mathbf{U}_{R,L} = \exp(\pm 2\pi i/3)\mathbf{U}_{R,L}$. Crucially, simple symmetry arguments (see Supplemental Material for details [23 23] and references [24, 25, 26]) can be used to demonstrate that both the gyromagnetic perturbation of the gap-filling medium shown in Fig.2(b), and the tripod-like perturbation of the rod's shape shown in Fig.2(c), are diagonal in the CP basis. Because these perturbations do not produce any TE/TM coupling, the perturbed electromagnetic solutions and their eigenfrequencies $\Omega_K^{e,m}(\delta \mathbf{k})$ are obtained by solving the following equation: $(\mathcal{H}_{0K}^{e,m} + \mathcal{H}_{1K}^{e,m})\mathbf{U} = \Omega_K^{e,m}\mathbf{U}$, where $\mathcal{H}_{1K}^{e,m}$ is the perturbation Hamiltonian.

Specifically, $\mathcal{H}_{1K}^{e,m} \equiv \mathcal{H}_{T,K}^{e,m} = \omega_D \Delta_{T,K}^{e,m} \hat{\sigma}_z$ for the former, and $\mathcal{H}_{1K}^{e,m} \equiv \mathcal{H}_{P,K}^{e,m} = \omega_D \Delta_{P,K}^{e,m} \hat{\sigma}_z$ for the latter, where the perturbation matrix elements are given by

$$\Delta_{T,K}^{e(m)} = -1/2 \int_{\Delta V} \left( \mathbf{h}_{e(m),\perp}^{R*} \cdot \Delta \bar{\bar{\mu}}_\perp \cdot \mathbf{h}_{e(m),\perp}^R - \mathbf{h}_{e(m),\perp}^{L*} \cdot \Delta \bar{\bar{\mu}}_\perp \cdot \mathbf{h}_{e(m),\perp}^L \right) dV, \quad (3.1)$$

where the field profiles are evaluated at $\delta \mathbf{k} = \mathbf{0}$ (e.g., $\mathbf{h}_e^R \equiv \mathbf{h}_e^{R, \mathbf{k}_\perp = K}$), and the volume $\Delta V$ of the two gaps is filled with a gyromagnetic material and

$$\Delta_{P,K}^{e(m)} = -1/2 \int_{\Delta V} \left( \mathbf{e}_{e(m)}^{R*} \cdot \mathbf{e}_{e(m)}^R - \mathbf{h}_{e(m)}^{R*} \cdot \mathbf{h}_{e(m)}^R - \mathbf{e}_{e(m)}^{L*} \cdot \mathbf{e}_{e(m)}^L + \mathbf{h}_{e(m)}^{L*} \cdot \mathbf{h}_{e(m)}^L \right) dV, (3.2)$$

where $\Delta V$ is the extruded volume of the tripods whose orientation determines the sign and the magnitude of $\Delta_{P,K}^{e(m)}$. For example, $\Delta_{P,K}^{e(m)} > 0$ for the tripod's orientation shown in Fig.2(c), but vanishes upon a $\pi/6$ rotation, and changes its sign upon a $\pi/3$ rotation [22]. While the above overlap integrals are, in general, different for the TE and TM coefficients, it is possible to design a photonic structure that satisfies $\Delta_{T,K}^e = \Delta_{T,K}^m \equiv \Delta_T(\mathbf{K})$ and $\Delta_{P,K}^e = \Delta_{P,K}^m \equiv \Delta_P(\mathbf{K})$, as we have done for the specific structures shown in Fig.2.

The complete decoupling between the TE/TM modes under the above perturbations is mathematically expressed by expanding the basis states from separate two-component vectors $\mathbf{U}_K^{e,m} = [a_{e,m}^R; a_{e,m}^L]$ to a four-component vector $\mathbf{V}_K = \mathbf{M}[\mathbf{U}_K^e; \mathbf{U}_K^m]$, where $\mathbf{M}$ is an arbitrary unitary $4 \times 4$ polarization-coupling matrix that does not mix the orbital states while coupling the TE/TM states. We will use $\mathbf{M} = \frac{1}{\sqrt{2}} \begin{pmatrix} 1 & 1 \\ 1 & -1 \end{pmatrix} \otimes \begin{pmatrix} 1 & 0 \\ 0 & 1 \end{pmatrix}$ that transforms from the TE/TM basis to the spin-up/spin-down (↑/↓) basis. The significance of the spin basis [4, 5, 8] is that it diagonalize the bi-anisotropic perturbation shown in Fig.2(a), which, unlike the tripod and gyromagnetic perturbations, directly couples the TE and TM states. In thus expanded spin-orbital basis, the effective perturbed Hamiltonians $\mathcal{H}_{1K}$ in the $K$ valley assume the following form: $\mathcal{H}_{T,K} = \omega_D \Delta_T \hat{s}_0 \hat{\sigma}_z$ for the gyromagnetic perturbation and $\mathcal{H}_{P,K} = \omega_D \Delta_p \hat{s}_0 \hat{\sigma}_z$ for the tripod perturbation. Here the Kronecker product (e.g., $\hat{s}_0 \hat{\sigma}_z \equiv \hat{s}_0 \otimes \hat{\sigma}_z$) of $2 \times 2$ matrices is used, and $\hat{s}_0$ and $\hat{s}_{x,y,z}$ are the unity and Pauli matrices operating on the space of spin states. As was demonstrated earlier [8] the perturbed Hamiltonian for the bi-anisotropic structure emulating the spin-orbit coupling is given by $\mathcal{H}_{SOC,K} = \omega_D \Delta_{SOC} \hat{s}_z \hat{\sigma}_z$, where the inter-mode coupling is given by $\Delta_{SOC} = -1/2 \int_{\Delta V} (e_e^{R*} \cdot e_m^R - h_e^{R*} \cdot h_m^R - e_e^{L*} \cdot e_m^L + h_e^{L*} \cdot h_m^L) \, dV$, where $\Delta V$ is the volume of metal inserted into one of the two gaps, and the sign of $\Delta_{SOC}$ is determined by which of the two rod-to-plate gaps is filled.

Finally, this calculation can be generalized to include both the $K$ and $K'$ valleys of the BZ by introducing an 8-component spinor, $\boldsymbol{\Psi} = [\boldsymbol{V}_K; \boldsymbol{T}\boldsymbol{V}_{K'}]$, where the transformation matrix $\boldsymbol{T}$ swaps the RCP and LCP orbital states. By introducing the Pauli matrices $\hat{\tau}_{x,y,z}$ and $\hat{\tau}_0$ operating on the valley subspace and using symmetry considerations, the effective $8 \times 8$ Hamiltonian spanning the orbit, valley, and spin subspaces can be generalized (see Supplemental Materials [23] for details) to $\mathcal{H}(\delta \boldsymbol{k}) = \mathcal{H}_0 + \mathcal{H}_{SOC/T/P}$, where $\mathcal{H}_0 = v_D(\delta k_x \hat{\tau}_z \hat{s}_0 \hat{\sigma}_x + \delta k_y \hat{\tau}_0 \hat{s}_0 \hat{\sigma}_y)$ describes the unperturbed PhG, and

$$\mathcal{H}_{SOC} = \omega_D \Delta_{SOC} \hat{\tau}_z \hat{s}_z \hat{\sigma}_z, \quad \mathcal{H}_T = \omega_D \Delta_T \hat{\tau}_z \hat{s}_0 \hat{\sigma}_z, \quad \mathcal{H}_P = \omega_D \Delta_P \hat{\tau}_0 \hat{s}_0 \hat{\sigma}_z \qquad (4)$$

are the perturbed Hamiltonians of the photonic structures shown in Figs.2(a-c), respectively. Equation (4) reveals that these three PTIs are formal photonic counterparts to electronic QSH, QH, and QVH topological insulators, respectively. The topological nature of the propagating spinors $\boldsymbol{\Psi}(\delta \boldsymbol{k})$ satisfying $\mathcal{H}\boldsymbol{\Psi} = \Omega \boldsymbol{\Psi}$ is captured by calculating the appropriate topological indices [30, 31, 32, 33] for each of the three photonic structures.

Specifically, the nonzero spin-Chern and valley-Chern indices exist even in photonic structures with $T$-symmetry that are known to have vanishing total Chern number, and can be very useful provided that inter-valley and spin-flipping transitions are suppressed [33]. The significance of these additional topological indices for photonics is that they can provide topological protection to surface waves between PTIs with opposing topological indices without breaking the time-reversal symmetry. Therefore, we calculate the local spin-valley indices [34] $C_{s,v} = \int_{BZ(v)} d^2 \delta \boldsymbol{k} \, [\boldsymbol{\nabla}_{\delta \boldsymbol{k}} \times \boldsymbol{A}(\delta \boldsymbol{k})]_z / 2\pi$, where $s = \uparrow, \downarrow$ is the spin state label, $v = K, K'$ is the valley label, and $BZ(v)$ is half of the BZ corresponding to $k_x > 0 (< 0)$ for $v = K(K')$, respectively. Here the local Berry connection [35, 36, 37] is $\boldsymbol{A}(\delta \boldsymbol{k}) = -i \boldsymbol{\psi}_v^{s\dagger}(\delta \boldsymbol{k}) \cdot \boldsymbol{\nabla}_k \boldsymbol{\psi}_v^s(\delta \boldsymbol{k})$, where $\boldsymbol{\psi}_v^s(\delta \boldsymbol{k})$ is a projection onto the $(s, v)$ spin-valley subspace of the full spinor $\boldsymbol{\Psi}(\delta \boldsymbol{k})$ propagating below the bandgap. The values of these indices for the Hamiltonians given by Eq.(4) are well known [30, 31, 32, 33, 34, 27], and here we reiterate them in the context of the three photonic structures shown in Fig.2: $2C_{\uparrow/\downarrow,v}^{SOC} = \pm 1 \times \text{sgn}(\Delta_{SOC})$, $2C_{s,v}^T = \text{sgn}(\Delta_T)$, and $2C_{s,K/K'}^P = \pm 1 \times \text{sgn}(\Delta_P)$. As expected, only the $T$-symmetry breaking gyromagnetic structure possesses a non-vanishing full Chern number obtained by summing over all spin and valley states. The numerically calculated propagation bands for the three types of PTIs are plotted in Figs.2(d-f) and marked by their corresponding topological indices. Note that while the topological gyromagnetically-induced bandgap has been demonstrated earlier [27, 28, 29] for a single TM mode, the structure shown in Fig.2(b) is unique in that both TE and TM polarizations are equally affected by the magnetic field, thereby emulating a spin-independent QH effect in electronic systems.

The existence of the local topological indices is crucial because they enable topologically protected surface waves (TPSWs) between heterogeneous PTIs. TPSWs emerge because of the impossibility of a continuous interpolation between the band structures across the interface separating two topological phases characterized by different topological indices. The number of such surface waves is obtained using the *bulk-boundary* correspondence principle [ 38] as the difference $\Delta C$ between topological indices of the top and bottom PTI claddings. For example, considering a QSH/QVH interface between PTIs with $\Delta_{SOC} > 0$ and $\Delta_P < 0$, we find that $\Delta C = C_{\uparrow,K}^{SOC} - C_{\uparrow,K}^{P} = 1$ and $\Delta C = C_{\downarrow,K'}^{SOC} - C_{\downarrow,K'}^{P} = -1$ for $K$ and $K'$ valley states, respectively. Therefore, one spin-up (down) forward (backward)-moving TPSW corresponding to the $K(K')$ valley is supported. This perfect "locking" between spin and valley DOFs arises because no topologically protected spin-down state in the $K$ valley can exist owing to $C_{\downarrow,K}^{SOC} - C_{\downarrow,K}^{P} = 0$. This is in contrast with the earlier studied [ 8] TPSWs at the homogeneous interface between two QSH-like PTIs with opposite signs of the $\Delta_{SOC}$ (see SM and Fig.S2 for review), where both spin-states exist in either valley.

The first-principles electromagnetic COMSOL simulation of the PBSs of the supercell shown in the inset of Fig.3(a) indeed reveals two counter-propagating (reciprocal) surface modes shown as blue solid lines. Both modes have negative refractive indices, i.e. their phase and group velocities point in the opposite directions. In the absence of inter-valley scattering, such surface modes are topologically protected against a broad class of lattice perturbations that preserve the $C_3$ point group symmetry and spin degeneracy [ 4, 5, 8]. An example of a non-reciprocal zigzag-shaped QSH/QH interface is shown in Fig.3(a), where only the forward-propagating TPSWs (dashed line: positive, solid line: negative refractive index) are supported because for both valleys $\Delta C = 1$ for the spin-up (↑) state. Below we demonstrate how, by combining these three types of the interfaces between photonic topological phases into an integrated heterogeneous PTI network, different valley and spin degrees of freedom can be spatially separated and filtered with nearly-perfect efficiency and without reflections.

One of the most intriguing topological effects in condensed matter physics is the possibility of "sorting" various degrees of freedom (valley, spin, helicity) by applying external fields. For example, valley DOFs can be spatially separated by taking advantage of different Berry curvatures $\Omega_{K(K')} = [\nabla_{\delta k} \times A(\delta k)]_z$ experienced by the electrons in the two valleys [ 15]. It turns out that similar sorting is accomplished for TPSWs when heterogeneous PTIs are interfaced with each other. An example of such photonic system is shown in Fig.4(a), where QVH and QH PTIs cavities are embedded as defects inside a QSH PTI. When a TPSW launched along the interface between two QSH PTI with opposite signs of $\Delta_{SOC}$ encounters a QVH PTI cavity as shown in Figs.4(b,c), it undergoes a valley-dependent deflection because the topological index difference $\Delta C = 1$ along the propagation path. Therefore, the two valley DOFs are spatially separated, with the $K(K')$-valley TPSW propagating along the upper (lower) path.

Note that valley-selective topologically protected photon transport shown in Fig.4(b,c) modeled using COMSOL is possible due to the conservation of the valley DOF. It can be rigorously demonstrated (see SM for details) that when the defect's border follows along the zigzag trajectory, the inter-valley scattering is identically zero, and full spatial separation of the $K(K')$-valley TPSWs is enabled. Therefore, the junction point at which the heterogeneous PTIs (QVH with $\Delta_P < 0$, QSH with $\Delta_{SOC} < 0$, and QSH with $\Delta_{SOC} > 0$) converge form an ultra-compact birefringent Y-junction that routes the photons according to their valley DOF. Such junctions may one day find use in quantum communications networks and other optical devices that rely on the entanglement between photons' polarization states. Moreover, if finite loss is introduced along one of the paths, then valley-filtering of the initial mixed-valley TPSW can be achieved.

The inclusion of a non-reciprocal QH PTIs into a heterogeneous PTI photonic network enables the development of ultra-compact photonic optical isolators. For example, by employing a QH PTI cavity shown in Fig.4(a) with finite loss in the lower arm, one can eliminate the back-propagating surface wave by routing it along the lossy path as shown in Fig.4(e). On the contrary, the forward-propagating wave routed along the upper arm as shown in Fig.4(d) does not experience losses, thus providing optical isolation capability beyond what has been achieved by earlier non-reciprocal photonic devices [39, 40]. The operational bandwidth of those designs is often limited by the need to optimize the coupling between reciprocal and nonreciprocal waveguides [40] and to avoid reflections. On the contrary, reflections in heterogeneous PTI networks are identically suppressed across the entire bandgap because of the spin conservation.

In conclusion, we have identified and designed three types of photonic topological insulators that support quantum spin-Hall, Hall, and valley-Hall topological phases. Such heterogeneous PTIs can be combined into integrated PTI networks that carry topologically protected electromagnetic surface waves, and can be used in a variety of photonic applications that rely on scattering-free wave propagation: optical insulators, delay lines, polarization splitters. Fundamentally, heterogeneous PTIs will enable the emulation of exotic interfaces between electronic topological phases that have been so far challenging to experimentally realize in naturally occurring materials. While scattering-free topological valley transport between two QVH topological insulators with opposite valley Chern numbers has been experimentally realized [20], no topologically protected edge states between heterogeneous TIs (e.g., QVH and QSH) have ever been observed in condensed matter systems. The three PTIs introduced in this Letter provide unique experimental platforms for emulating such heterogeneous interfaces using photons. By introducing nonlinear elements into these photonic platforms, fundamental questions such as the robustness of topological protection to many-body interactions will be experimentally addressed.

This work was supported by the National Science Foundation (NSF) Award PHY-1415547 and the Air Force Office of Scientific Research grant number FA9550-15-1-0075.

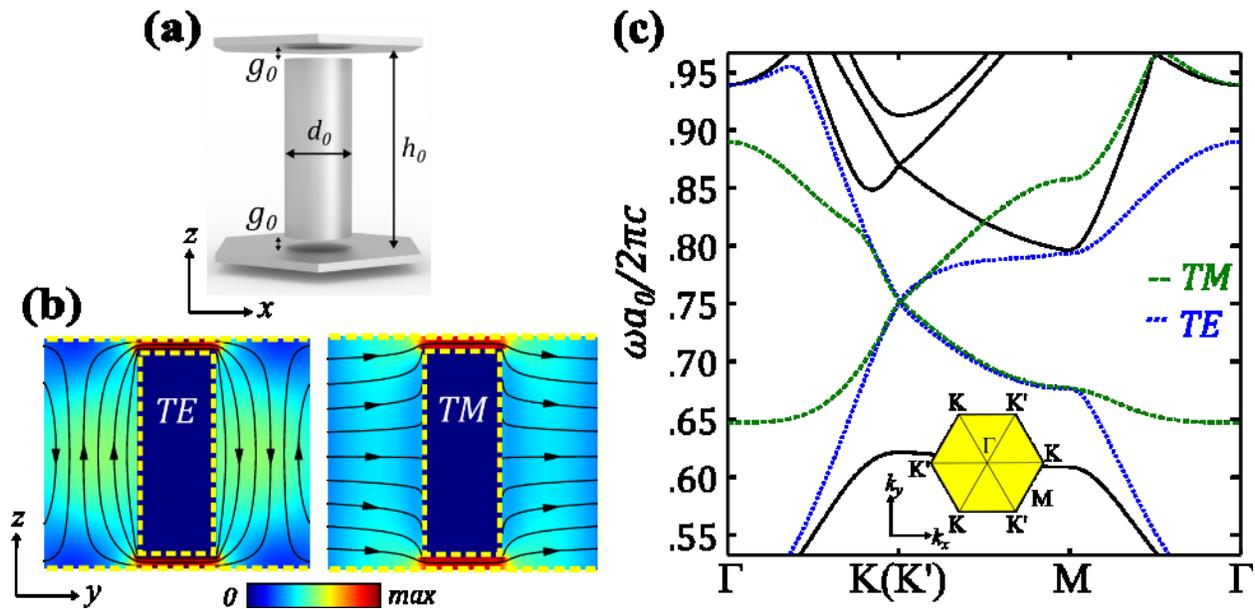

**Figure 1: The unperturbed "photonic graphene" (PhG) structure used for emulating photon equivalents of the spin and valley degrees of freedom. (a)** The unit cell of the PhG: metal rods arranged as a hexagonal array lattice with the lattice constant $a_0$. **(b)** Magnetic field profiles of the TE and TM modes at the $K$ point. **(c)** The PBS with TE and TM modes forming doubly-degenerate Dirac cones at $K(K')$ points. Design parameters: $h_0 = a_0$, $d_0 = 0.345a_0$, and $g_0 = 0.05a_0$.

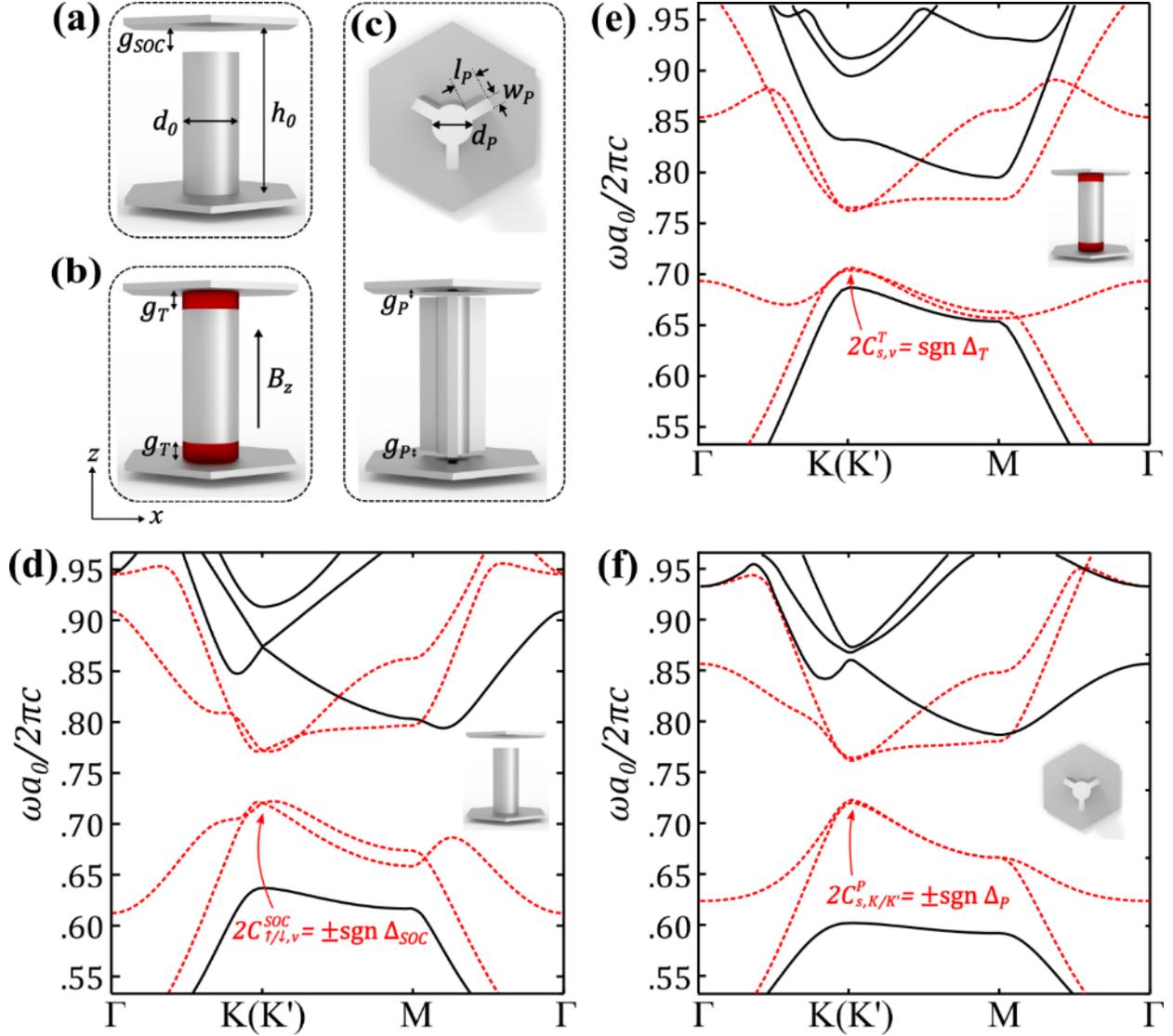

**Figure 2: A library of topological phases of light. (a,b,c)** The unit cells of the three PTIs emulating QSH, QH, and QVH effects, respectively. In (b), the out-of-plane magnetic field is applied to a gyromagnetic material (red disks) with the following constitutive parameters: $\epsilon = 1$, $\mu_{xx/yy/zz} = 1$, $\mu_{xy} = -\mu_{yx} = -0.8i$, and $\mu_{ij} = 0$ otherwise. **(d,e,f)** The band structures corresponding to the PTIs in **(a,b,c)**, respectively. with the bandgaps induced by various gapping method. The bands at the $K(K')$ points are doubly-degenerate. The local spin-valley Chern indices of the lower bands are listed. PTIs' parameters, $h_0 = a_0$, $d_0 = 0.345a_0$, $g_{em} = 0.15a_0$, $g_T = 0.1a_0$, $g_P = 0.03a_0$, $d_P = 0.2a_0$, $l_P = 0.116a_0$, and $w_P = 0.06a_0$ are chosen to ensure spin-degeneracy and approximately equal band gaps.

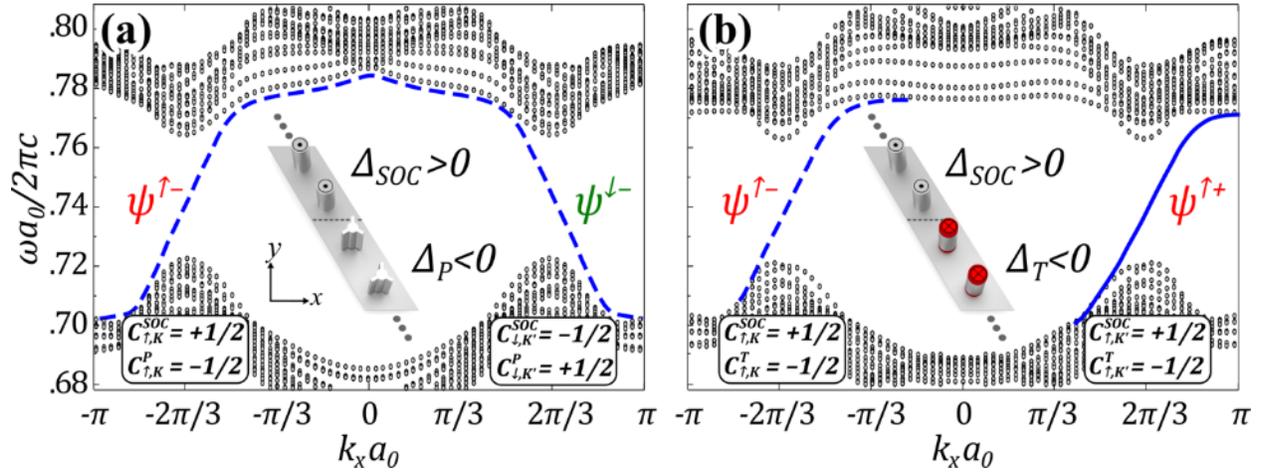

**Figure 3: Topologically protected surface waves along zigzag interfaces between heterogeneous PTI claddings.** Photonic band structures of the super-cells containing **(a)** QSH/QVH, and **(b)** QSH/QH PTI interfaces. The super-cells contain a single cell along the propagation *x*-direction and 20 cells on each side of the interface. Black circles: bulk modes, blue solid/dashed lines: dispersion curves of the TPSWs with positive/negative refractive index, ↑/↓ labels the spin. Boxed tags: spin-Chern and valley-Chern topological indices of the bulk modes below the bandgap that belong to top/bottom claddings.

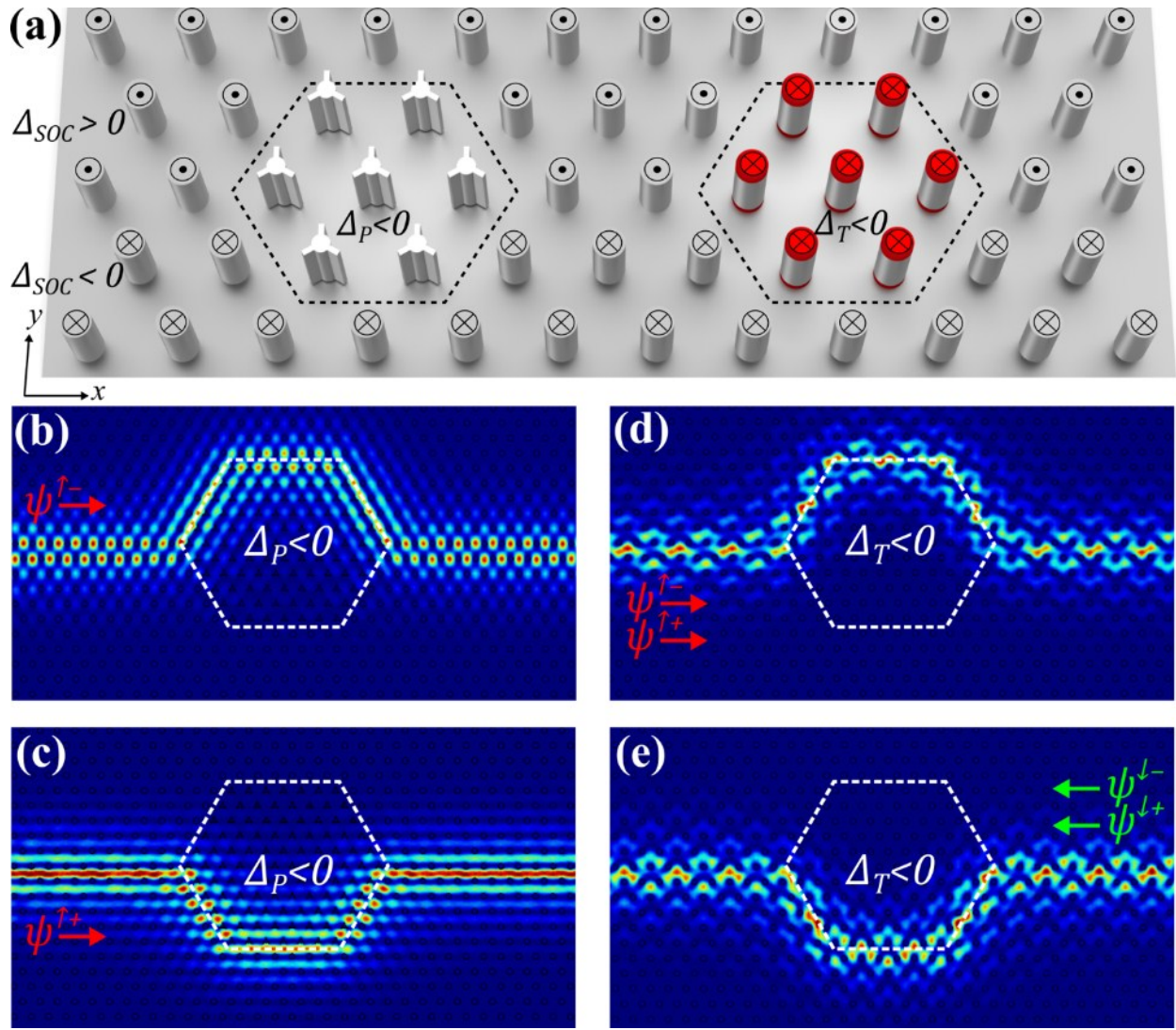

**Figure 4: Reciprocal and non-reciprocal heterogeneous photonic topological networks. (a)** Schematic: QVH- and QH-PTI defect cavities embedded in a QSH PTI matrix containing an interface between $\Delta_{SOC} > 0$ (dots) and $\Delta_{SOC} < 0$ (crosses) QSH PTIs. **(b,c)** Valley-dependent deflection of TPSWs: **(b)** $K'$-valley (negative index) TPSWs are deflected to the upper arm, and **(c)** $K$-valley (positive index) TPSWs are deflected to the lower arm of an embedded QVH-PTI cavity. **(d,e)** Spin-dependent (nonreciprocal) deflection of TPSWs: **(d)** spin-↑ (left-to-right propagating) TPSWs are deflected to the upper arm, and **(e)** spin-↓ (right-to-left propagating) TPSWs are deflected to the lower arm of an embedded QH-PTI cavity. Colors: $|E_z \pm H_z|^2$ for ↑/↓ spin states, respectively.

# Supplemental Material

*Perturbation theory:*

The effective Hamiltonian description of the photonic topological insulators (PTIs) in the main texts is based on the perturbation theory applied to an extended electromagnetic cavity for estimating the coupling strength between four lowest-order dipolar modes in the PTIs. It is sufficient to consider only these four modes, because in the frequency range (about 7% photonic bandwidth) the PTIs are operated only these four modes are present, and other higher-order modes are far away (more than 7% away from the centered operating frequency). There are in general two types of perturbations in an electromagnetic cavity; one can perturb the cavity either by deforming the metallic boundary of the cavity or by varying the material properties of the cavity fillings. Here we need both. In our PTIs, we introduce coupling between the modes by adding/removing small piece of metal around the rods or by filling the gaps with gyromagnetic materials.

For the first type of the perturbations with deformation of metallic boundary, the change of the eigenfrequencies of the modes can be expressed as an overlap integral of the fields in the unperturbed cavity according to Slater's theory [1]. In perturbation regime, the electric and magnetic fields of the perturbed modes, $(\boldsymbol{E'}, \boldsymbol{H'})$, can be expanded as

$$\boldsymbol{E'} = \sum_n a_n \boldsymbol{E}_n \text{ and } \boldsymbol{H'} = \sum_n a_n \boldsymbol{H}_n, \tag{S1}$$

where $(\boldsymbol{E}_n, \boldsymbol{H}_n)$ are the eigenfields of the unperturbed cavity forming an orthonormal basis, and $a_n$'s represent the degree of hybridization between the unperturbed modes. Following the generalization of the Slater's theory [2], which considers the condition where the unperturbed modes are not necessary far separated in frequency and thereby the hybridization between the modes could be important, the matrix equation can be derived as:

$$\begin{pmatrix} \omega_{11} + \kappa_{11} & \kappa_{12} & \cdots \\ \kappa_{21} & \omega_{22} + \kappa_{22} & \cdots \\ \vdots & \vdots & \ddots \end{pmatrix} \begin{pmatrix} a_1 \\ a_2 \\ \vdots \end{pmatrix} = \omega' \begin{pmatrix} a_1 \\ a_2 \\ \vdots \end{pmatrix} \tag{S2}$$

where $\omega_{mn}$, with $m, n = 1, 2, 3 \cdots$, is the eigenfrequency of the corresponding unperturbed modes; the frequency of the perturbed modes, $\omega'$'s, are obtained by solving the eigenvalue problem of Eq.(S2); the coupling coefficient, $\kappa_{mn}$, is given as an overlap integral:

$$\kappa_{mn} = -\int_{\Delta V} (\omega_m \boldsymbol{E}_m^* \cdot \boldsymbol{E}_n - \omega_n \boldsymbol{H}_m^* \cdot \boldsymbol{H}_n) \, dV \tag{S3}$$

where $\Delta V$ is the perturbed volume where an extra piece of metal is insert, and the eigenfields are normalized as $\int_V (|\boldsymbol{E}_n|^2 + |\boldsymbol{H}_n|^2) \, dV = 1$ with $V$ being the volume of a unitcell. In our PTIs, we are interested in the hybridization between four dipolar modes degenerate at the Dirac frequency, $\omega_D$, as shown in the main texts. Eq.(S3) can be thereby further simplified as $\kappa_{mn} = -\omega_D \int_{\Delta V} (\boldsymbol{E}_m^* \cdot \boldsymbol{E}_n - \boldsymbol{H}_m^* \cdot \boldsymbol{H}_n) \, dV$, and we can defined an unitless coupling strength as

$$\Delta_{mn} \equiv \frac{\kappa_{mn}}{\omega_D} = -\int_{\Delta V}(\boldsymbol{E}_m^* \cdot \boldsymbol{E}_n - \boldsymbol{H}_m^* \cdot \boldsymbol{H}_n)\, dV \tag{S4}$$

In the case of perturbing a cavity by changing material properties, the basic formulation remain unchanged except Eq.(S4) is modified according to the standard cavity perturbation theory [3].

$$\Delta_{mn} = -\int_{\Delta V}(\boldsymbol{E}_m^* \cdot \Delta\bar{\bar{\epsilon}} \cdot \boldsymbol{E}_n + \boldsymbol{H}_m^* \cdot \Delta\bar{\bar{\mu}} \cdot \boldsymbol{H}_n)\, dV \tag{S5}$$

where $\Delta\bar{\bar{\epsilon}}$ and $\Delta\bar{\bar{\mu}}$ is the changing permittivity and permeability and $\Delta V$ is the region where the material properties are changed. In the next section, we apply the perturbation theory and specifically calculate the coupling strength in the case of PTIs.

*Properties of the circularly-polarized basis, and of the perturbations of the Photonic Graphene structure*

In this section, we demonstrate that both the non-reciprocal (breaking the $T$-symmetry) gyromagnetic perturbation of the gap-filling medium shown in Fig.2(b), and the of the reciprocal tripod-like perturbation of the rod's shape (breaking the $P$-symmetry) shown in Fig.2(c) of this letter, are diagonal in the orbital basis comprised of circularly polarized (CP) states of right-hand circular polarized (RCP) and left-hand circular polarized (LCP) photons at the $K$-point of the Brillouin zone (BZ). Specifically, we demonstrate that the perturbed Hamiltonians associated with the above perturbations are $\mathcal{H}_T^{e(m)} = \Gamma_T^{e(m)}\hat{\sigma}_0 + \omega_D \Delta_T^{e(m)} \hat{\sigma}_z$ for the former, and $\mathcal{H}_P^{e(m)} = \Gamma_P^{e(m)}\hat{\sigma}_0 + \omega_D \Delta_P^{e(m)} \hat{\sigma}_z$ for the latter. Here $\hat{\sigma}_0$ is a unit matrix, and $\hat{\sigma}_z$ is a Pauli matrix, both in the CP basis. The overall frequency offsets $\Gamma_T^{e(m)}$ and $\Gamma_P^{e(m)}$ are not particularly important, but the frequency splitting amplitudes $\Delta_T^{e(m)}$ and $\Delta_P^{e(m)}$ will be considered below in some detail. These two amplitudes can be recast as $\Delta_T^{e(m)} = \Delta_{RR}^{T,e(m)} - \Delta_{LL}^{T,e(m)}$ and $\Delta_P^{e(m)} = \Delta_{RR}^{P,e(m)} - \Delta_{LL}^{P,e(m)}$, where the diagonal perturbation amplitudes, from Eqs.(S4,S5), are given by $\Delta_{nn}^{T,e(m)} = -\int_{\Delta V}[\boldsymbol{h}_{\perp,e}^{n*} \cdot \Delta\bar{\bar{\mu}} \cdot \boldsymbol{h}_{\perp,e}^n]\, dV$ and $\Delta_{nn}^{P,e(m)} = -\int_{\Delta V}[\boldsymbol{e}_{e(m)}^{n*} \cdot \boldsymbol{e}_{e(m)}^n - \boldsymbol{h}_{e(m)}^{n*} \cdot \boldsymbol{h}_{e(m)}^n]\, dV$, and $n = L, R$ is the polarization state label.

To understand why (i) $\Delta_{T,P}^{e(m)}$ are non-vanishing, and (ii) $\Delta_{ij}^{T,e(m)} = \Delta_{ij}^{P,e(m)} = 0$ if $i \neq j$, we first consider the profiles of the electromagnetic fields $\boldsymbol{e}_{e(m)}^R$ and $\boldsymbol{h}_{e(m)}^R$ pertaining to the RCP and LCP expansion basis members of the unperturbed "photonic graphene" (PhG) structure. The fields are evaluated at the $K$-point of the BZ of the hexagonal lattice of the PhG. The unit cell of the PhG is shown in Fig.S1(a). The orbital states are defined by their transformation with respect to the $2\pi/3$ rotation operation $\mathcal{R}_3$: $\mathcal{R}_3 \boldsymbol{e}_{e(m)}^R = \eta \boldsymbol{e}_{e(m)}^R$ and $\mathcal{R}_3 \boldsymbol{e}_{e(m)}^L = \eta^* \boldsymbol{e}_{e(m)}^L$, where $\eta = \exp(i2\pi/3)$. The same relations hold for the magnetic fields: $\mathcal{R}_3 \boldsymbol{h}_{e(m)}^R = \eta \boldsymbol{h}_{e(m)}^R$ and $\mathcal{R}_3 \boldsymbol{h}_{e(m)}^L = \eta^* \boldsymbol{h}_{e(m)}^L$.

Useful illustration of the properties of the RCP and LCP states is provided by the Langrangian density $\mathcal{L}_R^{e(m)}(\mathbf{r}_\perp) = |\mathbf{e}_{e(m)}^R|^2 - |\mathbf{h}_{e(m)}^R|^2$ of the RCP state, and $\mathcal{L}_L^{e(m)}(\mathbf{r}_\perp) = |\mathbf{e}_{e(m)}^L|^2 - |\mathbf{h}_{e(m)}^L|^2$ of the LCP state in the symmetry plane $z = 0$. We plot $\mathcal{L}_L^m(\mathbf{r}_\perp)$ and $\mathcal{L}_R^m(\mathbf{r}_\perp)$ in Figs.S1(b,c), respectively, for the TM modes, and $\mathcal{L}_L^e(\mathbf{r}_\perp)$ and $\mathcal{L}_R^e(\mathbf{r}_\perp)$ in Figs.S1(d,e), respectively, for the TE modes. Because these fields are the integrant of the perturbation matrix elements such as $\Delta_{RR}^{P,e(m)}$ and $\Delta_{LL}^{P,e(m)}$, one can anticipate/estimate the values of these matrix elements by observing the value of the overlap field in the perturbed volume $\Delta V$. It can be seen from Figs.S1(b-e) that the overlap integrals $\Delta_{RR}^{P,e(m)}$ and $\Delta_{LL}^{P,e(m)}$ strongly depend on the orientation of the perturbed tripod. For example, for the tripod orientation shown in Fig.2(c), the tripod's arms strongly overlap with the LCP state, but not with the RCP state. Therefore, it is apparent through visual appreciate that $\Delta_{RR}^{P,e(m)} \neq \Delta_{LL}^{P,e(m)}$ and $\Delta_{RR}^{T,e(m)} \neq \Delta_{LL}^{T,e(m)}$ for at least some orientations of the tripod. Also, because $\mathcal{L}_R^{e(m)}(\mathbf{r}_\perp)$ and $\mathcal{L}_L^{e(m)}(\mathbf{r}_\perp)$ are the mirror images of each other with respect to the x-axis, an inverted tripod would flip the signs of $\Delta_P^{e(m)}$.

Helicity densities defined as $\mathcal{G}_R^{e(m)}(\mathbf{r}_\perp) = i\hat{z} \cdot [\mathbf{h}_{e(m)}^{R*} \times \mathbf{h}_{e(m)}^R] + c.c.$ for the RCP, and $\mathcal{G}_L^{e(m)}(\mathbf{r}_\perp) = i\hat{z} \cdot [\mathbf{h}_{e(m)}^{L*} \times \mathbf{h}_{e(m)}^L] + c.c.$ for the LCP orbital states, are evaluated at $z = g_0/2$ (middle of the gap region) and plotted in Figs.S1(f-i) for the TE and TM wave polarizations. The integrals of the helicity densities over the gap between the rod and the plate are directly proportional to the relevant matrix elements $\Delta_{RR}^{T,e(m)}$ and $\Delta_{LL}^{T,e(m)}$ through the non-reciprocal proportionality coefficient $i\Delta\mu_{xy} = -i\Delta\mu_{yx} = \kappa$. From the plots in Fig.S1(f-i) we observe that $\Delta_{LL}^{T,e(m)} = -\Delta_{RR}^{T,e(m)}$. They as well indicate that by flipping the applied magnetic field along z-axis, one can flip the sign of $\Delta_{LL}^{T,e(m)}$ and $\Delta_{RR}^{T,e(m)}$ because $\kappa \propto B_z$.

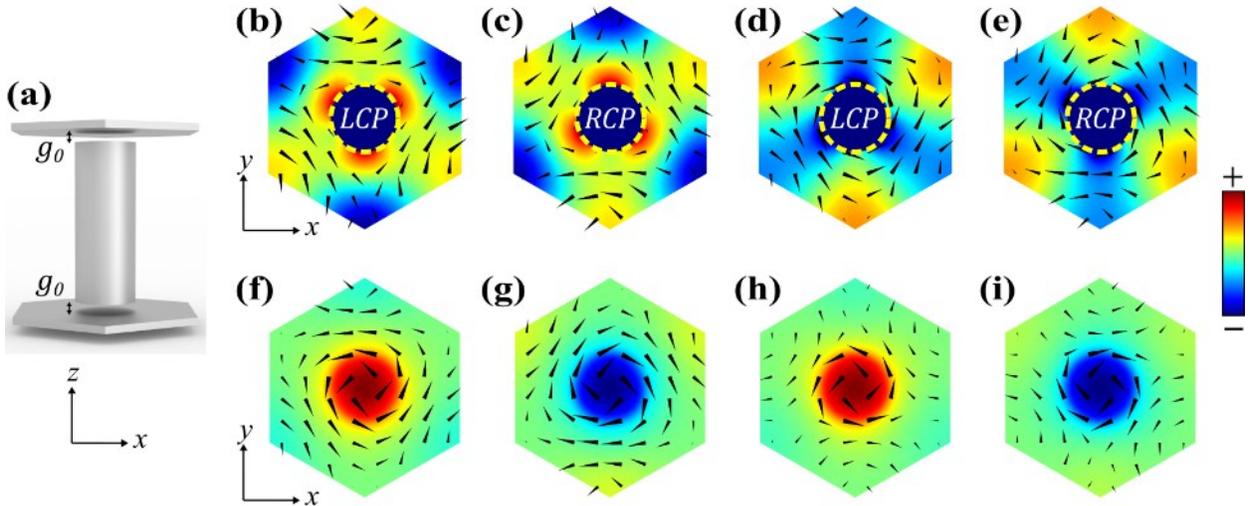

**Figure S1: Overlap fields of the unperturbed PhG (a)** Schematic of the unitcell of the PhG with symmetric gaps. **(b,c)**: the color-coded TM overlap fields (Lagrangian density) $\mathcal{L}_L^m(\mathbf{r}_\perp, z = 0) = |\mathbf{e}_m^L|^2 - |\mathbf{h}_m^L|^2$ for the LCP, and $\mathcal{L}_R^m(\mathbf{r}_\perp, z = 0) = |\mathbf{e}_m^R|^2 - |\mathbf{h}_m^R|^2$ for the RCP basis modes, respectively. **(d,e)**: same as **(b,c)**, but for the TE polarization: plots of $\mathcal{L}_L^e(\mathbf{r}_\perp, z = 0) = |\mathbf{e}_e^L|^2 - |\mathbf{h}_e^L|^2$ for the LCP, and $\mathcal{L}_R^e(\mathbf{r}_\perp, z = 0) = |\mathbf{e}_e^R|^2 - |\mathbf{h}_e^R|^2$ for the RCP basis modes, respectively. Arrows: in-plane power flux illustrating the orbital state of the modes. **(f-i)**: same as **(b-**

e), but the plotted color-coded quantities are the helicity densities $\mathcal{G}_{R(L)}^{e(m)}(\mathbf{r}_\perp, z = g/2) = i\hat{z} \cdot \left[\mathbf{h}_{e(m)}^{R(L)*} \times \mathbf{h}_{e(m)}^{R(L)}\right] +$ c.c. for all possible orbital and polarization states: $\mathcal{G}_L^m(\mathbf{r}_\perp, z = g/2)$ for the TM-polarized LCP modes in **(f)**, $\mathcal{G}_R^m(\mathbf{r}_\perp, z = g/2)$ for the TM-polarized RCP modes in **(g)**, $\mathcal{G}_L^e(\mathbf{r}_\perp, z = g/2)$ for the TE-polarized LCP modes in **(h)**, and $\mathcal{G}_R^e(\mathbf{r}_\perp, z = g/2)$ for the TM-polarized RCP modes in **(i)**. Arrows: in-plane power flux illustrating the orbital state of the modes. Physical dimensions of the PhG lattice: $h_0 = a_0$, $d_0 = 0.345a_0$, $g = 0.05a_0$.

## Coupling between RCP and LCP modes

In this section we use simple symmetry arguments to demonstrate that both the gyromagnetic perturbation of the gap-filling medium shown in Fig.2(b), and the tripod-like perturbation of the rod's shape shown in Fig.2(c), are diagonal in the CP basis. The symmetry argument is based on the fact that the RCP/LCP states, and the two perturbations (the former breaks the $T$- and the latter breaks the $P$-symmetry) possess the $C_3$ wave vector symmetry. One can see that, by performing a $2\pi/3$ rotation ($\mathcal{R}_3$) around the z-axis: the perturbed structures are not altered. Below, we prove that the cross overlap integrals between the RCP and LCP must be zero under these perturbations.

We start with tripod perturbation, which has the off-diagonal elements in the perturbation matrix like $\Delta_{RL}^{P,e(m)} = -\int_{\Delta V}\left[\mathbf{e}_{e(m)}^{R*} \cdot \mathbf{e}_{e(m)}^{L} - \mathbf{h}_{e(m)}^{R*} \cdot \mathbf{h}_{e(m)}^{L}\right] dV$. Because the system is unchanged under $\mathcal{R}_3$, the following equality must satisfied:

$$\Delta_{RL}^{P,e(m)} = \mathcal{R}_3 \Delta_{RL}^{P,e(m)} \tag{S6}$$

The explicit operation of $\mathcal{R}_3 \equiv \mathcal{R}_{\theta=2\pi/3}$ contents two parts: (i) mapping the argument of a function such that $(x,y,z) \to (x',y',z')$ where $(x',y',z')^T = \bar{\bar{R}}_{-\theta}(x,y,z)^T$; $\bar{\bar{R}}_\theta = \begin{pmatrix} \cos\theta & -\sin\theta & 0 \\ \sin\theta & \cos\theta & 0 \\ 0 & 0 & 1 \end{pmatrix}$ and $\theta = 2\pi/3$; (ii) rotating the components of a vector such that $(v_x, v_y, v_z)^T \to \bar{\bar{R}}_\theta (v_x, v_y, v_z)^T$. So together one has the rotation operation of a vector field:

$$\mathcal{R}_\theta \begin{pmatrix} v_x(x,y,z) \\ v_y(x,y,z) \\ v_z(x,y,z) \end{pmatrix} = \bar{\bar{R}}_\theta \begin{pmatrix} v_x(x',y',z') \\ v_y(x',y',z') \\ v_z(x',y',z') \end{pmatrix} \tag{S7}$$

From the definition of the $R$ and $L$ fields, we know that $\mathcal{R}_3 \mathbf{e}_{e(m)}^R = \eta \mathbf{e}_{e(m)}^R$ and $\mathcal{R}_3 \mathbf{e}_{e(m)}^L = \eta^* \mathbf{e}_{e(m)}^L$, where $\eta = \exp(i2\pi/3)$. Thus, Eq.(S6) becomes

$$\Delta_{RL}^{P,e(m)} = \int_{\Delta V'}\left[\mathcal{R}_3\mathbf{e}_{e(m)}^{R*} \cdot \mathcal{R}_3\mathbf{e}_{e(m)}^{L} - \mathcal{R}_3\mathbf{h}_{e(m)}^{R*} \cdot \mathcal{R}_3\mathbf{h}_{e(m)}^{L}\right] dV$$

$$= \eta^{*2} \int_{\Delta V}\left[\mathbf{e}_{e(m)}^{R*} \cdot \mathbf{e}_{e(m)}^{L} - \mathbf{h}_{e(m)}^{R*} \cdot \mathbf{h}_{e(m)}^{L}\right] dV$$

$$= \eta^{*2} \Delta_{RL}^{P,e(m)}$$

To satisfy the above equality, $\Delta_{RL}^{P,e(m)}$ has to be zero.

For the gyromagnetic perturbation, one has the off-diagonal elements in the perturbation matrix like $\Delta_{RL}^{T,e(m)} = -\int_{\Delta V}\left[\boldsymbol{h}_{\perp,e(m)}^{R*} \cdot \Delta\bar{\bar{\mu}} \cdot \boldsymbol{h}_{\perp,e(m)}^{L}\right]dV$. Because the system is unchanged under $\mathcal{R}_3$, the following equality similar to Eq.(S6) must satisfied:

$$\Delta_{RL}^{T,e(m)} = \mathcal{R}_3 \Delta_{RL}^{T,e(m)} \tag{S8}$$

Because of the present of $\Delta\bar{\bar{\mu}}$, it is not immediately clear how can one reduce Eq.(S8) like we did for the case of tripod perturbation. However from the fact that the magnetic field is applied along z-axis, one has the only non-vanishing elements in $\Delta\bar{\bar{\mu}}$ to be $\mu_{xy} = -\mu_{yx}$, and thereby $\Delta\bar{\bar{\mu}}$ can be written as a scalar function multiplied by a rotational matrix such that

$$\Delta\bar{\bar{\mu}} = \mu_{xy}(x,y,z)\bar{\bar{R}}_{-\pi/2} \tag{S9}$$

Because the 2D rotational matrix like $\bar{\bar{R}}_\theta$ are commute with each other, we have

$$\begin{aligned}
&\mathcal{R}_3\left[\Delta\bar{\bar{\mu}}(x,y,z) \cdot \boldsymbol{h}_{\perp,e(m)}^{L}(x,y,z)\right] \\
&= \bar{\bar{R}}_{2\pi/3}\left[\mu_{xy}(x',y',z')\bar{\bar{R}}_{-\pi/2}\boldsymbol{h}_{\perp,e(m)}^{L}(x',y',z')\right] \\
&= \left[\mu_{xy}(x',y',z')\bar{\bar{R}}_{-\pi/2}\bar{\bar{R}}_{2\pi/3}\boldsymbol{h}_{\perp,e(m)}^{L}(x',y',z')\right] \\
&= \left[\mu_{xy}(x,y,z)\bar{\bar{R}}_{-\pi/2}\eta^*\boldsymbol{h}_{\perp,e(m)}^{L}(x,y,z)\right] \\
&= \eta^*\Delta\bar{\bar{\mu}}(x,y,z) \cdot \boldsymbol{h}_{\perp,e(m)}^{L}(x,y,z)
\end{aligned}$$

(S10)

The last equality in Eq.(S10) uses the properties of $\mu_{xy}(x,y,z)$ to be of $\mathcal{R}_3$ rotational invariant. Using this result, we can now reduce Eq.(S8),

$$\begin{aligned}
\Delta_{RL}^{T,e(m)} &= \int_{\Delta V'}\left[\mathcal{R}_3\boldsymbol{h}_{\perp,e(m)}^{R*} \cdot \mathcal{R}_3\left(\Delta\bar{\bar{\mu}} \cdot \boldsymbol{h}_{\perp,e(m)}^{L}\right)\right]dV \\
&= \int_{\Delta V}\left[\eta^*\boldsymbol{h}_{\perp,e(m)}^{R*} \cdot \eta^*\left(\Delta\bar{\bar{\mu}} \cdot \boldsymbol{h}_{\perp,e(m)}^{L}\right)\right]dV \\
&= \eta^{*2}\Delta_{RL}^{T,e(m)}
\end{aligned}$$

Therefore to satisfy the above equality, $\Delta_{RL}^{T,e(m)}$ has to be zero.

*The Effective Hamiltonians for the Designed Perturbations:*

In the main texts, the effective Hamiltonian approach is adopted for describing the hybridization of the modes in PhG under various perturbations and also for further calculation of the spin-valley Chern indices. The effective Hamiltonian has the form: $\mathcal{H}(\delta \mathbf{k}) = \mathcal{H}_0 + \mathcal{H}_{SOC/T/P}$ with the unperturbed term: $\mathcal{H}_0 = v_D(\delta k_x \hat{t}_z \hat{s}_0 \hat{\sigma}_x + \delta k_y \hat{t}_0 \hat{s}_0 \hat{\sigma}_y)$, and the perturbed term:

$$\mathcal{H}_{SOC} = \omega_D \Delta_{SOC} \hat{t}_z \hat{s}_z \hat{\sigma}_z, \quad \mathcal{H}_T = \omega_D \Delta_T \hat{t}_z \hat{s}_0 \hat{\sigma}_z, \quad \mathcal{H}_P = \omega_D \Delta_P \hat{t}_0 \hat{s}_0 \hat{\sigma}_z \tag{S11}$$

Here we relist Eq.(4) in Eq.(S11) for the convenience. It has been shown that the effective Hamiltonian approach can be used to describe an unperturbed PhG [4], and the extension of this description for incorporating spin degree of freedom (DOF) has been demonstrated in refs.[5, 6] so that $\mathcal{H}_0$ takes its form. The detail derivation of $\mathcal{H}_{SOC}$ has also been shown in ref.[6], and it has been proved that $\mathcal{H}_0 + \mathcal{H}_{SOC}$ is identical to the Kane-Mele's Hamiltonian [7]. In the section, we explain how the other two perturbed Hamiltonians, $\mathcal{H}_T$ and $\mathcal{H}_P$, have their form as shown in Eq.(S11).

As mentioned, both the tripod-like perturbation and the gyromagnetic perturbation reduced the wavevector symmetry group from $C_{3v}$ to $C_3$ and therefore opening a bandgap at the $K(K')$ point. This effect of gapping the Dirac point by breaking the inversion or time-reversal symmetry can be understood from the symmetry properties of the RCP and LCP modes. In the main texts, we define the CP modes to have the field symmetry satisfying that $\mathcal{R}_3 V_{K(K')}^{R/L} = \exp(\pm 2\pi i/3) V_{K(K')}^{R/L}$. In the unperturbed PhG as shown in Fig.S1, the two CP modes are connected by the mirror operation ($\mathcal{M}$) with respect to $\Gamma - K(K')$ direction such that $\mathcal{M} V_{K(K')}^{R/L} = V_{K(K')}^{L/R}$. There are two important symmetry operations that relate the modes at two valleys: $T$-symmetry ($\mathcal{T}$) and $P$-symmetry ($\mathcal{P}$) operations such that $\mathcal{T} V_K^{R/L} = V_{K'}^{L/R}$ and $\mathcal{P} V_K^{R/L} = V_{K'}^{R/L}$. Note that $\mathcal{P}$ is equivalent to an $\pi/2$ rotation along z-axis under which the direction of energy flow remains the same. From $T$-symmetry operation, It becomes clear now why the state vector of $K'$ valley has $R$ and $L$ labeled in a reverse order such that the total state vector is defined as $\mathbf{\Psi} = [V_K; \mathbf{T} V_{K'}]$ with the transformation matrix $\mathbf{T}$ swapping the RCP and LCP orbital states. It is well-known that a time-reversible photonic system has symmetric band structure, and the states at $\mathbf{k}$ and $-\mathbf{k}$ in BZ form a Kramer pair with the same frequency; for this reason, it is natural to use the convention so that their state vectors stay invariant under time-reversal. One may also notice that the three symmetry operations introduced so far are related by $\mathcal{M} = \mathcal{P}\mathcal{T}$. From this, it is conclusive that without breaking $P$-symmetry and $T$-symmetry, $\mathcal{M}$ brings the system back to itself, and the RCP and LCP states connected by $\mathcal{M}$ are degenerate at the $K(K')$ point; breaking either $T$-symmetry or $P$-symmetry thus lifts the Dirac degeneracy.

With this insight about the symmetry properties of the states, we can now write down the explicit matrix representation of the perturbed Hamiltonian, $\mathcal{H}_T$ and $\mathcal{H}_P$. For the perturbation which breaks $P$-symmetry only, one have the degeneracy of the RCP and LCP states lifted at the $K(K')$ valleys. The perturbation we considered has $C_3$ point group symmetry which respects the symmetry of the RCP and LCP states, thereby does not couple the two states (also see the previous section for quantitative explanation), and $\mathcal{H}_P$ should be diagonalized in the orbit subspace. Moreover the perturbation preserves $T$-symmetry such that the Kramer pairs remains degenerate in frequency. Therefore by combining these two symmetry properties, $\mathcal{H}_P = \omega_D \Delta_P \hat{t}_0 \hat{s}_0 \hat{\sigma}_z$. Note that the spin subspace is trivially introduced because the TE and TM modes do not couple; an arbitrary linear superposition for constructing this subspace is allowed, and it has only trivial effect on the perturbed Hamiltonian.

For the perturbation which breaks $T$-symmetry only, one still has $C_3$ point group symmetry so that the corresponding matrix representation, $\mathcal{H}_T$, should also be diagonalized in the orbit subspace while lifting the degeneracy of the $R$ and $L$ states. However since the $P$-symmetry is preserved, an $\pi/2$ rotation (equivalent to the parity operation) along z-axis brings the system back to itself. This operation also takes $V_K^{R/L} \to V_{K'}^{R/L}$, and thereby the perturbation matrix can only has the form: $\mathcal{H}_T = \omega_D \Delta_T \hat{\tau}_z \hat{s}_0 \hat{\sigma}_z$ for preserving the necessary degeneracies.

*Topologically protected surface waves along zigzag interfaces between heterogeneous PTI claddings.*

In the main texts, we demonstrate that the spin-valley indices can be used to predict the existence of the topologically protected surface waves (TPSWs) based on the bulk-boundary correspondence principle [ 8]. For the sake of completeness, we also show the simulated band structures of the surface modes for the other three interfacing between heterogeneous PTI claddings in Fig.S2 and Fig.S3.

On an QSH/QSH interface as shown in Fig.S2(a), there are four TPSWs because $\Delta C_K = C_{\uparrow(\downarrow),K}^{SOC1} - C_{\uparrow(\downarrow),K}^{SOC2} = \pm 1$, and $\Delta C_{K'} = C_{\uparrow(\downarrow),K'}^{SOC1} - C_{\uparrow(\downarrow),K'}^{SOC2} = \pm 1$. The field profiles of the four TPSWs labeled by their spins and refractive indices are shown in Fig.S2(c,d) with the corresponding photonic band structure (PBS) as shown in Fig.S2(b).

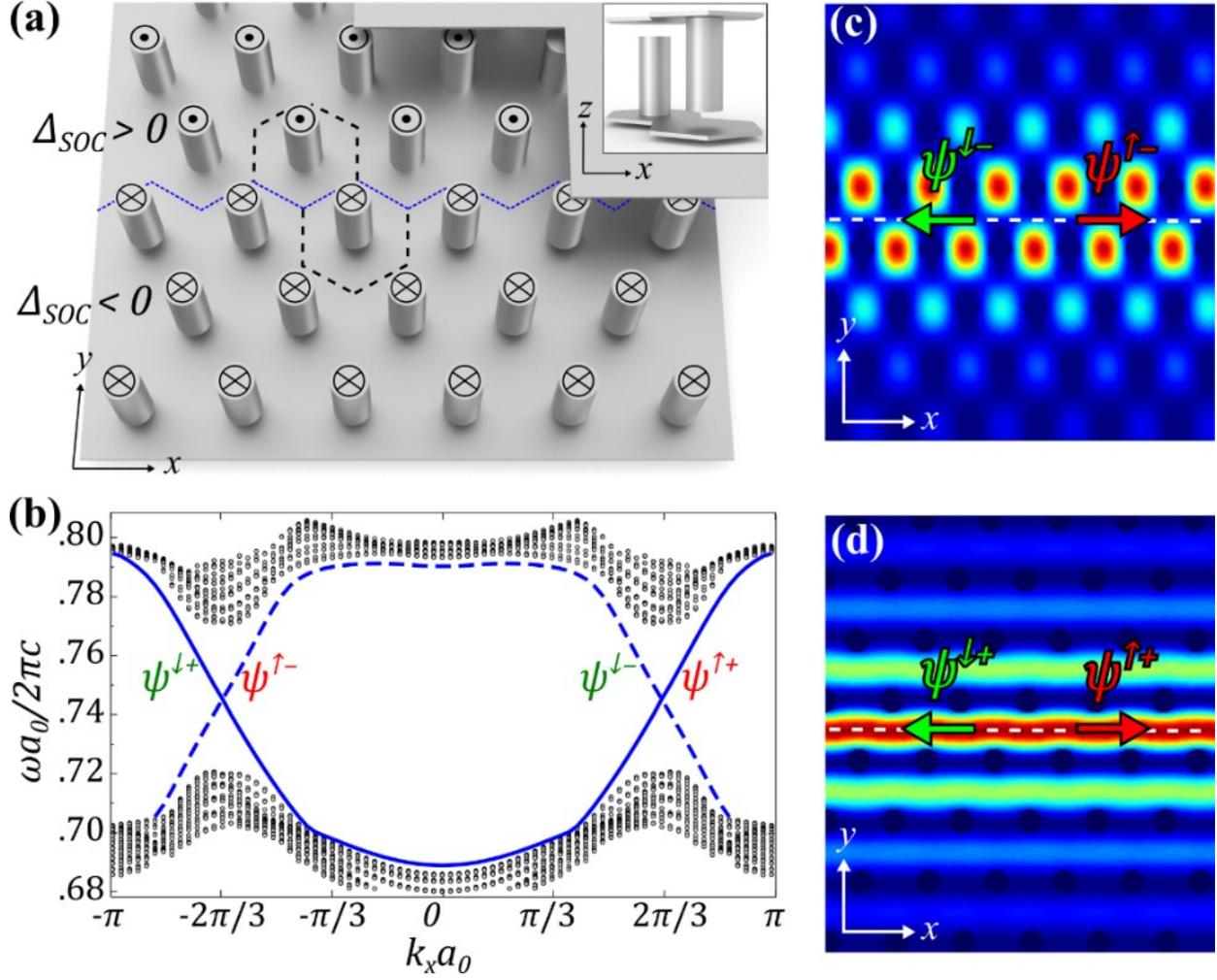

**Figure S2: Topologically protected edge modes along QSH/QSH zigzag interface. (a)** Top views of the zigzag interface (thin dashed line) between two QSH PTIs introduced in Fig.2(a). Thick dashed line outlines the side view (inset) of two adjacent unit cells on the edge. Black arrows on rods indicate $\text{sgn}(\Delta_{SOC})$. **(b)** PBS of a supercell (single cell along *x*-direction, 30 cells on each side of the interface). Black circles: bulk modes, blue solid/dashed lines: dispersion curves of the TPSWs with positive/negative refractive index. ↑/↓ labels spin, and +/− labels refractive index. **(c,d)** Field profiles of the TPSWs with positive/negative refractive index respectively. Color: $|E_z \pm H_z|^2$ for ↑/↓ respectively. Red and green arrows show the propagation direction of the edge mode with different spins.

Fig.S3 shows the same combinations of interfacing heterogeneous PTIs as Fig.3 but with the opposite signs of $\Delta_P$ and $\Delta_T$. The effects of changing sign of the perturbation on the TPSWs are straightforward from the bulk-boundary correspondence. One can see in Fig.S3(a), the allowed TPSWs become those with positive refractive index, and in Fig.S3(b), the allowed TPSWs are now spin-down.

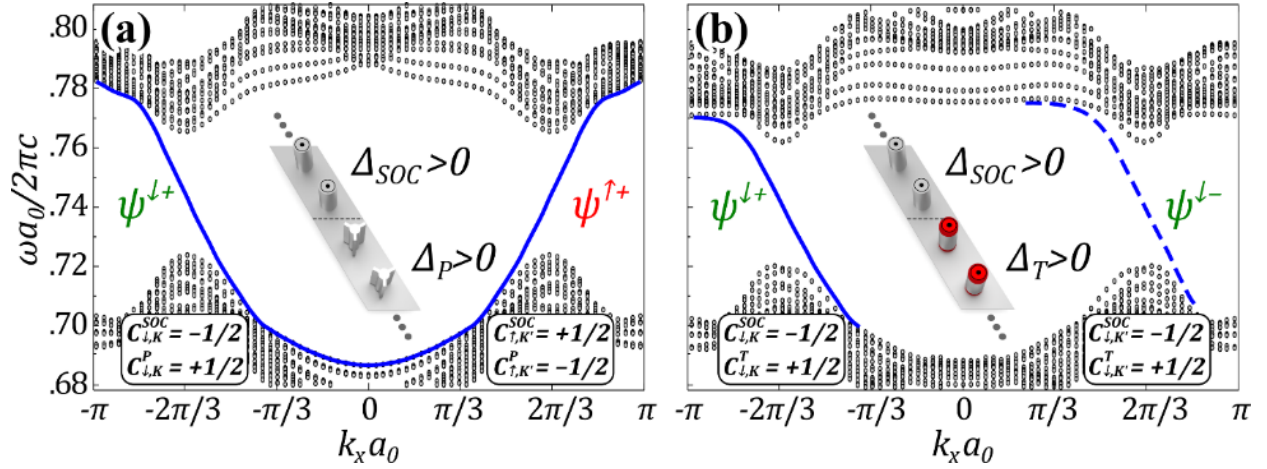

**Figure S3: Topologically protected surface waves along zigzag interfaces between heterogeneous PTI claddings.** PBSs of the super-cells containing **(a)** QSH/QVH, and **(b)** QSH/QH PTI interfaces. The super-cells contain a single cell along the propagation *x*-direction and 20 cells on each side of the interface. Black circles: bulk modes, blue solid/dashed lines: dispersion curves of the TPSWs with negative/positive refractive index, ↑/↓ labels the spin. Boxed tags: spin-Chern and valley-Chern topological indices of the bulk modes below the bandgap that belong to top/bottom claddings.

*Suppression of Inter-Valley Scattering and Conservation of the Valley Degree of Freedom under 'zigzag' Perturbation of the Interface between PTIs with Different Topological Valley Indices*

In this section, we analytically explain under which type of perturbations the valley DOF is conserved. In other words, once a surface wave is launched at certain valley on an interface, while it is propagating, the scattering only takes place within the valley in an effective 2D BZ of the whole integrated system. Although the integrated system as shown in Fig.4(a) is not, strictly speaking, a periodic structure, one can consider those surface state has local field profiles near the edges the same as the propagating bulk states in an unperturbed PhG near the Dirac points. The only difference between the bulk and edge states is that, as we engineer the perturbation, the edge states only propagate along the direction where the edges are present, and the propagation paths are engineered depending on which lattice sites are perturbed. Thus to understand the scattering properties between the edge states along different edges, studying the field overlaps of the unperturbed PhG is sufficient as long as the PTIs are still in perturbation regime.

Because the valley conservation in our discussion does not involve spin DOF, we can consider only one representative field component of the eigenmodes, say $E_z(\bm{r}_\perp, z = h/2) = \psi_{K(K')}(\bm{r})$ for the TM modes (The TE modes follows the identical argument below), where $\bm{r} = (x, y)$. At Dirac points of the $K$ and $K'$ valleys, $\psi_{K(K')}(\bm{r})$ can be expressed in the Bloch form:

$$\psi_K(\bm{r}) = u_K(\bm{r})e^{i\bm{K}\cdot\bm{r}} \tag{S12}$$

$$\psi_{K'}(\bm{r}) = u_{K'}(\bm{r})e^{i\bm{K'}\cdot\bm{r}}, \tag{S13}$$

where $u_K(r)$ and $u_{K'}(r)$ are functions with the periodicity of the lattice; $K = e_x\, 4\pi/3a_0$ and $K' = -e_x\, 4\pi/3a_0$; $r = (x, y)$. The overlapped field of the same valley (intra-valley), the $K$ valley, is then

$$\psi_K^* \psi_K = u_K^* u_K e^{i(K-K)\cdot r} = u_K^* u_K. \tag{S14}$$

Since $u_K^* u_K$ has the periodicity identical to the lattice, one can expect that the perturbation sitting on some lattice sites (with the same finite volume at the center of a unit cell) are always going to add up. This is because the value of the overlap integral of $\psi_K^* \psi_K$ over perturbation volume in every unit cell is exactly the same. However this is not the case for the inter-valley (between two different valleys) one. We shall see that for the inter-valley overlapped field $\psi_{K'}^* \psi_K$, the perturbations on the nearby sites tend to cancel with each other. The inter-valley overlapped field has in fact different periodicity from that of the lattice:

$$\begin{aligned}
\psi_{K'}^* \psi_K &= e^{i(K-K')\cdot r} u_{K'}^* u_K \\
&= e^{i\frac{2}{3}(b_1+b_2)\cdot r} \sum_{m,n} a_{mn}^{KK'}\, e^{i(mb_1+nb_2)\cdot r} \\
&= \sum_{m,n} a_{mn}^{KK'}\, e^{i[(3m+2)b_1' + (3n+2)b_2']\cdot r} \\
&= \sum_{m',n'} b_{m'n'}^{KK'}\, e^{i(m'b_1' + n'b_2')\cdot r}
\end{aligned}$$

(S15)

where $b_{1,2}$ is the reciprocal lattice vectors, and $b_{1,2}' = 1/3\, b_{1,2}$, $m' = 3m + 2$, $n' = 3n + 2$. The last line of Eq.(S15) shows that $\psi_{K'}^* \psi_K$ has the original hexagonal symmetry, but has the period changed to $3a_0$ characterized by a new set of reciprocal vectors $b_1'$ and $b_2'$. With this mental picture of the inter-valley overlapped fields, we derive the special condition of the perturbation that gives zero overlap integral of $\psi_{K'}^* \psi_K$ (i.e. the perturbation that conserves the valley DOF).

Consider again the perturbations sitting on some lattice sites with the same finite volume. The difference, this time, is that the overlap integral corresponding to each site is no longer identical. The value of the integral varies from site to site with period $3a_0$ along 6 special directions ($l\pi/6$ with $l = 0, 1, \ldots, 5$) which is known as the directions of zigzag edge or the direction of the $K$ and $K'$ points. If we place the perturbations on 3 lattice sites in series along the directions of zigzag edge, the overlap integral reads

$$\int_{\Delta V = \Delta V_1 + \Delta V_2 + \Delta V_3} dV\, \psi_{K'}^* \psi_K$$

$$
\begin{aligned}
&= \int_{\lambda=0}^{3a_0} \int_{\lambda_\perp=-\infty}^{\infty} d\lambda \, d\lambda_\perp \, w(\lambda_\perp) \, h \sum_{l'} \Lambda_{l'} e^{il'\frac{2\pi}{a_0}\hat{\lambda}\cdot r} \sum_{m',n'} b^{KK'}_{m'n'} e^{i(m'b'_1+n'b'_2)\cdot r} \\
&= \Delta A \sum_{l',m',n'} \Lambda_{l'} b^{KK'}_{m'n'} \int_{\lambda=0}^{3a_0} d\lambda \, e^{i\left(l'\frac{2\pi}{a_0}\hat{\lambda}+m'b'_1+n'b'_2\right)\cdot r} \\
&= \Delta A \sum_{l',m',n'} \Lambda_{l'} b^{KK'}_{m'n'} \frac{3ia_0}{2\pi} \cdot \frac{1 - e^{i2\pi[3l'+(m'+n')\cos\phi+1/\sqrt{3}(m'-n')\sin\phi]}}{3l' + (m'+n')\cos\phi + 1/\sqrt{3}(m'-n')\sin\phi} \\
&= \Delta A \sum_{l',m',n'} \Lambda_{l'} b^{KK'}_{m'n'} \frac{3ia_0(1 - e^{i2\pi I})}{2\pi I} = 0
\end{aligned}
$$

(S16)

where $\hat{\lambda} = [\cos\phi, \sin\phi]$ and $r = [\lambda\cos\phi, \lambda\sin\phi]$ with $\phi = l\pi/6$ and $l = 0, 1, \ldots, 5$; $w(\lambda_\perp)$: a localized function along $\lambda_\perp$ (the coordinate along the perpendicular direction of $\hat{\lambda}$), $h$: the height of the perturbation volume, and the Fourier series in the direction of $\hat{\lambda}$ make the integration region continuous; $\Delta A = \int_{\lambda_\perp=-\infty}^{\infty} d\lambda_\perp \, w(\lambda_\perp) \, h$ is the vertical cross section of the perturbation volume. The integer I in the last line of Eq. (S16) is

$$
I \equiv 3l' + (m'+n')\cos\phi + 1/\sqrt{3}(m'-n')\sin\phi = \begin{cases} 3l' \pm m' \pm n', & l = 0, 3 \\ 3l' \pm m', & l = 1, 4 \\ 3l' \mp n', & l = 2, 5 \end{cases} \quad (S17)
$$

As shown in Eq. (S17), I is an integer as long as $\phi$ is of that along the zigzag edge, and if so, the overlap integral [Eq. (S16)] is identically zero. This result shows that as long as the perturbations are designed to be the same at wherever lattice sites we put them and they perturb $3N$ lattice sites in series along the direction of zigzag edge, the inter-valley scattering is prohibited and the valley DOF is conserved. We refer this type of perturbations 'zigzag' as opposed to the other 'armchair' type of perturbation.

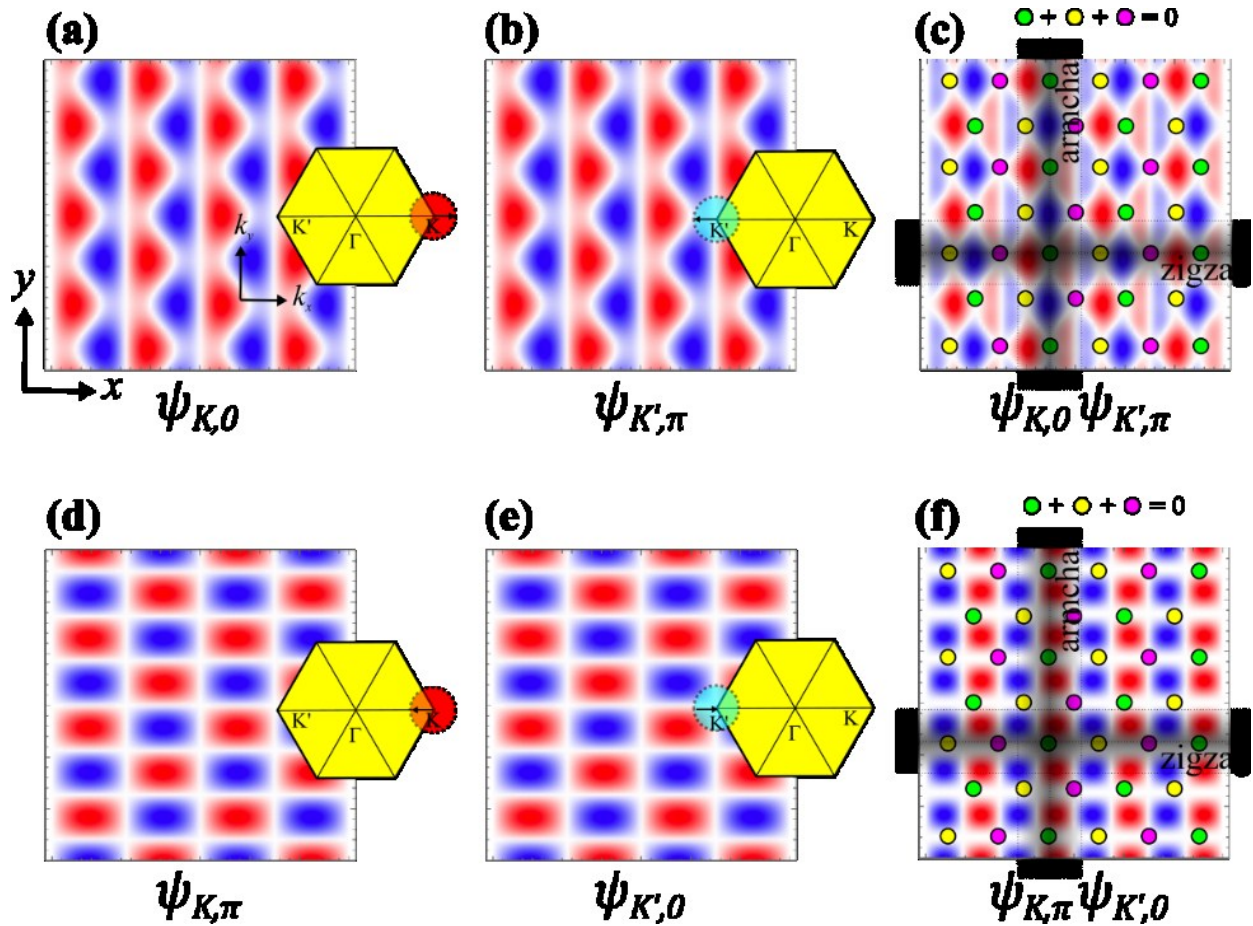

**Figure S4: Inter-valley overlap fields and the valley conservation.** **(a,b)** Field profiles of forward-moving $K$-valley and backward-moving $K'$-valley respectively. **(c)** The overlap field between (a) and (b); color dots distinct the different strength of coupling proportional to the overlap integral in the region near the lattice sites; Gray bands mark the direction of zigzag and armchair. **(d,e,f)** Same as (a,b,c) but for the case of backward-moving $K$-valley and forward-moving $K'$-valley.

Fig.S4 illustrates the idea of the valley conservation under zigzag perturbation. Figs.S4(a-c) show the overlap field between forward-moving $K$-valley and backward-moving $K'$-valley, and Figs.S4(d-f) show the overlap field between backward-moving $K$-valley and forward-moving $K'$-valley. The color dots mark the different strength of on-site perturbations which are directly proportional to the field overlap integral in the perturbation region. One can see that along zigzag direction the perturbations of different lattice sites tend to cancel each other as we analytically shown above, whereas they tend to add up along the armchair direction. That is, the armchair-type of perturbation in general do not conserve the valley DOF.